\documentclass[a4paper,aps, prb,amsmath,amssymb, superscriptaddress, showpacs, twocolumn, floatfix]{revtex4}
\usepackage{graphicx, braket, mathrsfs, nicefrac, epsfig, multirow, ulem, verbatim, subfigure}
\newcommand{\opcl}[1]{\left(#1\right)}
\newcommand{\ti}[1]{\tilde{#1}}
\newcommand{\avg}[1]{\left\langle#1\right\rangle}

\begin{document}
\title{Statistics of radiation at Josephson parametric resonance}
\author{Ciprian Padurariu}
\affiliation{Kavli Institute of Nanoscience, Delft University of Technology, Lorentzweg 1,
2628 CJ Delft, The Netherlands}
\author{Fabian Hassler}
\affiliation{Institute for Quantum Information, RWTH Aachen University,
D-52056 Aachen, Germany}
\author{Yuli V. Nazarov}
\affiliation{Kavli Institute of Nanoscience, Delft University of Technology, Lorentzweg 1,
2628 CJ Delft, The Netherlands}

\begin{abstract}

Motivated by recent experiments, we study theoretically the full counting statistics of radiation emitted below the threshold of parametric resonance in a Josephson junction circuit. In contrast to most optical systems, a significant part of emitted radiation can be collected and converted to an output signal. This permits studying the correlations of the radiation.  

To quantify the correlations, we derive a closed expression for full counting statistics in the limit of long measurement times. We demonstrate that the statistics can be interpreted in terms of uncorrelated bursts each encompassing $2N$ photons, this accounts for the bunching of the photon pairs produced in course of the parametric resonance.   We present the details of the burst rates. In addition, we  study the time correlations within the bursts and discuss experimental signatures of the statistics deriving the frequency-resolved cross-correlations.
\end{abstract}
\maketitle

%Introduction general
\section{Introduction}
Parametric resonance \cite{ParametricResonance} is one of the most fundamental
and frequently applied non-linear phenomena. If a non-linear oscillator with the resonant frequency $\Omega_0$ 
is a.c. driven  at frequency $2\Omega \approx 2 \Omega_0$, 
a coherent resonant response at frequency $\Omega$ emerges provided the driving amplitude exceeds an instability threshold set by the non-linear parameters of the oscillator. 
While the coherent response is absent below the threshold, the parametric resonance is manifested
there by enhanced fluctuations with frequencies close to $\Omega$.
%\cite{EnhancedNoise}.  
In the quantum realm ($\hbar
\Omega \gg k_B T$ with $T$ the temperature), these fluctuations  can be regarded as an emission of radiation.
An elementary radiation event is an emission of a pair of photons of the
frequency $\approx \Omega$ caused by absorption of a single photon of the
frequency $2\Omega$.  In quantum optics, the corresponding phenomenon is
called down-conversion\cite{DownConversion} since a single photon is converted
into two.  The down-conversion is a base of optical quantum-information
applications\cite{ReviewOpticsQuantumInfo}.  The phenomenon has been employed
to produce squeezed states of light\cite{SquezedStates} and pairs of
quantum-entangled photons\cite{FirstEntanglement}.

%Introduction specific

It seems natural to assume that the statistics of the radiation is that of
uncorrelated elementary events, each event being the emission of a
correlated/entangled pair.  In most optical experiments, this assumption is
correct and practical.  However, it relies on the fact that only a minor
fraction of emitted pairs is actually detected.  Detected events are separated
by large time intervals and thus do not show any correlation.  Recently,
a set of pioneering experiments \cite{Bunch} has advanced quantum non-linear optics
into the microwave frequency range.  Thereby, atoms are replaced by
superconducting qubits made using Josephson junctions, and the radiation is
confined to transmission lines and electrical oscillators.  The latter
represents a large technical advantage in comparison with an optical
experiment due to the fact the radiation is not lost and concentrated.  This
enhances the non-linearities of the system.

%Immediately relevant introduction

Very recently, accurate measurements of the radiation emitted by a dc
voltage-biased Josephson junction embedded in a microwave resonator have been
reported.  \cite{BrightSide} The Josephson generation frequency $\omega_J = 2
eV/\hbar$ can be tuned to double the resonant frequency, fulfilling the
conditions of parametric resonance.  Importantly, up to $50 \%$ of the emitted
radiation can be detected and the fluctuations of the detector signal can be
quantified as well.\cite{Private} This motivated us to study the statistics of
the radiation in this setup.
The full photon counting statistics of the degenerate optical parametric oscillator has been addressed in \cite{VyasSingh89} for a specific case when the driving frequency  is precisely $2 \Omega_0$. Since the observation of non-Poissonian features of these statistics requires collection efficiency not achievable in optical setups, this work has not attracted the attention it deserves. Let us note that the measurements of statistics do not require the detector to be an actual counter giving the output signal in terms of discrete numbers of counts. A continuous detector output would suffice to quantify cumulants of the radiation intensity fluctuations and thereby characterize the statistics. 

%Key paragraph

In this paper, we revisit the full counting statistics (FCS) of radiation below the instability threshold bringing this to the context of Josephson circuit.   
This regime is interesting since 
despite the fact that the field correlations are entirely Gaussian under these
conditions, the statistics are highly nontrivial. We restrict our attention to FCS in the limit of the long measurement times. We recover the results of  \cite{VyasSingh89} in a different conceptual framework that is directly based on the Keldysh-action treatment of dissipative Josephson dynamics. We extend the results to the case of an arbitrary 
mismatch between driving frequency and $2\Omega_0$. We provide  an interpretation of the
statistics. In this interpretation,  an elementary event is a radiation burst that
encompasses correlated emission of $N$ pairs, rather than an emission of a single pair.  
This is a manifestation of  
photon bunching. We outline the similarities with the results \cite{KindermannBeenakkerNazarov}concerning the bunching in a single-photon regime. The rate of $N$-burst does not diverge upon approaching the threshold.  However, larger $N$ are
favored closer to the threshold. This results in a divergence of the average
radiation intensity and its higher moments.  We support this interpretation by
investigating time correlations of the emission events.  Further, we quantify
the frequency-resolved fluctuations of the radiation. 
 The correlations of the spectral-resolved intensity permit a relatively easy experimental observation and we present several relevant formulas to facilitate those.

 The structure of the paper is as follows. We describe the setup in Section \ref{sec:setup}.
 We discuss the Keldysh-action of the setup and introduce the counting field required for computing the statistics in Section \ref{sec:action}. We evaluate the FCS in Section \ref{sec:statistics} and present the results for the photon Fano factor and big deviations from the equilibrium.
In Section \ref{sec:interpretation} we give the interpretation in terms of bursts computing the partial rates of corresponding $2k$-photon processes.
We discuss two limiting cases of the FCS described in Section \ref{sec:limit}.
We analyze the time-dependent fluctuations of radiation intensity in  Section \ref{sec:time} making use of the Keldysh propagator of the fields. In Section \ref{sec:resolved} we discuss the experimental significance of the frequency-resolved intensity correlations and quantify those. 
We conclude in Section \ref{sec:conclusion} and give details of the field propagator in the Appendix.

%And now concrete points

%1:Setup 
\section{Setup}
\label{sec:setup}
\begin{figure}
\centerline{\includegraphics[width=0.9\linewidth]{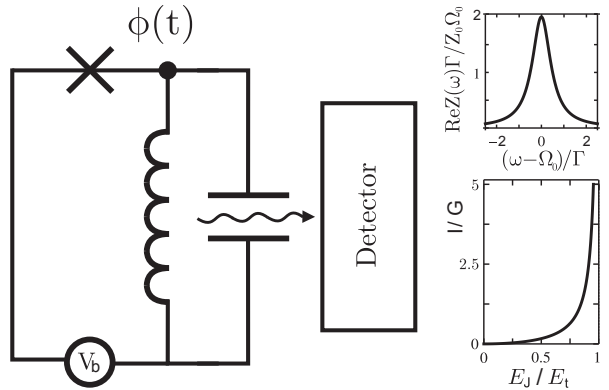}}
\caption{Setup. The Josephson junction with Josephson energy $E_J$ (cross in the Figure) is connected to a resonator represented with an inductor and a capacitor. All resonator losses are absorbed by a detector and converted to a measurable signal with efficiency $f$. Right pane up: the impedance of the resonator in the vicinity of the resonant frequency $\Omega_0$. Right pane down: the emission intensity from the resonator as function of  $E_J$, $E_t$ corresponds to instability threshold. }
\label{fig:setup}
\end{figure}

We concentrate on a setup similar to Ref. \onlinecite{BrightSide}. In main, it comprises  a Josephson junction biased by a d.c. voltage source that is connected to a high-quality (that is, quality factor $Q\gg 1$) microwave resonator (Fig. \ref{fig:setup}). We describe the resonator losses by the damping rate $\Gamma$. All photons leaving the resonator are absorbed by a (counting) detector. It is characterized by an efficiency $f$, a fraction of photons that are successfully counted. The impedance near the resonant frequency $\Omega_0$ reads
\begin{equation}
\label{eq:resonant-impedance}
Z(\omega \approx \Omega_0) = \frac{ Z_0 \Omega_0}{-i\nu + \Gamma/2}
\end{equation}
$\nu \equiv \omega -\Omega_0 \ll \Omega_0$ being the frequency mismatch. We will mostly concentrate on quantum limit of vanishing temperature $k_B T \ll \hbar \Omega$.  In this case, no photons come from the environment and the detector reading is the number of photons emitted from the resonator. The setup is characterized with a single quantum variable $\phi(t)$, related to the voltage across the inductor by means of Josephson relation $\dot{\phi} = 2e V(t)/\hbar$. The superconducting phase difference across the junction, $\phi_J$, is contributed by $\phi(t)$ and the voltage source, $\phi_J = \phi + 2 e V_b t$.

We will assume that the impedance far from the resonance, $Z_0$, is sufficiently small at the quantum scale, that is, 
$Z_0 G_Q \ll 1$, $G_Q \equiv e^2/\pi \hbar$. Under this assumption, the junction is effectively in a low-impedance environment, and the contributions to the quantum fluctuations of $\phi(t)$ coming from frequencies far from $\Omega_0$, $\delta \phi \simeq \sqrt{Z_0 G_Q}$, can be safely neglected. Since practical impedances are in the range of tens of Ohms, this assumption is well-justified. We stress that the assumption does not restrict the impedance near $\Omega_0$, $Z \simeq Z_0 (\Omega_0/\Gamma)$ that can exceed the quantum scale at sufficiently big quality factors. This however is not needed for our approach to be valid: we only require $Q\gg 1$.

%TALK ABOUT THE CORRESPONDENCE BETWEEN NUMBER OF PHOTONS AND COOPER PAIRS. 

\section{Keldysh action}
\label{sec:action}
Quantum dynamics of Josephson junction are well-explored \cite{Schoen}.
The most general and adequate quantum description of the setup is provided \cite{Schoen} by a Keldysh-type path-integral over the variables $\phi^{\pm}(t)$, $\pm$ refers to the c-values of the quantum variable $\phi(t)$ at the forward(backward) part of the Keldysh contour. 
The "partition
function" ${\cal Z}$ that is identically $1$ in the traditional Keldysh approach is given by the path integral over the configurations of $\phi^{\pm}(t)$ weighted with the factor  $e^{iS}$, $S$ being the quantum action expressed in Keldysh variables (from now on, we set $\hbar=1$)
\begin{equation}\label{eq:part_fun}
  {\cal Z}= \int \mathcal{D}[\phi^+(t)]\mathcal{D}[\phi^-(t)] e^{iS}
\end{equation}
The whole action is composed from Josephson and environmental part, $S= S_\text{env} + S_J$. 

The action of the Josephson junction is simply given by its energy $E(\phi_J) = - E_J \cos(\phi_J)$
and reads 
\begin{equation}\label{eq:actionj}
  S_J = -E_J \int \!dt \,[\cos \phi^+_J(t) - \cos \phi_J^-(t)]
\end{equation}
 The superconducting phase difference across the junction is contributed by the 
d.c. bias voltage $V_b$ applied, so we substitute 
\begin{equation}\label{eq:phase}
\phi_J (t)=\frac{2e}{\hbar}V_b\; t+\phi(t)
\end{equation}
into this part of the action.

The action of the environment for a general frequency-dependent impedance $Z(\omega)$ reads
\begin{equation}\label{eq:actionenv} 
S_\text{env} = \frac{i}{8\pi G_Q}
  \int
  \frac{d\omega}{2\pi} \sum_{\alpha,\beta =\pm} (\phi_\omega^\alpha)^*
  M_{\alpha\beta}(\omega) \phi_\omega ^\beta
\end{equation}
with $\phi^\alpha_\omega = \int\!dt\, e^{i \omega t} \phi^\alpha(t)$, and
\begin{align}
%\label{eq:y}
M(\omega) = & \omega\left( {\rm Im}Y(\omega) \left[\begin{array}{cc} 1& 0\cr 0 & -1 \end{array}\right]+ 
{\rm Re}Y(\omega) \left[\begin{array}{cc} 0& -1\cr 1 & 0 \end{array}\right]+ \right.\nonumber\\
 (2n(\omega) +1) &\left.{\rm Re}Y(\omega) \left[\begin{array}{cc} 1& -1\cr -1 & 1 \end{array}\right]\right).
\end{align}
where $n(\omega)=(\exp(\omega/k_BT)-1)^{-1}$ gives the Bose-Einstein filling factor at temperature $T$ and
the admittance $Y(\omega) \equiv Z^{-1}(\omega)$.
%%%\begin{equation}\label{eq:y}
%%%  M(\omega) = \omega\begin{pmatrix}
%%%     (n(\omega)+1) Z^{-1}(\omega)+ n(\omega)Z^{-1}(\omega)^* &  - 2n(\omega) \mathop{\rm Re}Z^{-1}(\omega) \\
%%%    - 2(n(\omega) +1)\mathop{\rm Re}Z^{-1}(\omega) & n(\omega) Z^{-1}(\omega)+(n(\omega)+1)  Z^{-1}(\omega)^*
%%%  \end{pmatrix}
%%%\end{equation}
%%%The environment impedance determines the voltage fluctuations over the junction.
%%%%
%%%\begin{equation}\label{eq:voltage}
%%%V(t)=V+\int^{\infty}_{-\infty}\frac{d\omega}{2\pi} (I_c \sin(\phi(t)))(\omega)Z(\omega)\; e^{-i\omega t}
%%%\end{equation}
%%%%
Variation of the action $S$ with respect to $\phi^{+} -\phi_{-}$ at $\phi^{+} \approx \phi^{-} \approx \phi$
reproduces the "classical'' equation of motion that disregards thermal and quantum fluctuations of $\phi$,
\begin{equation}
\int\!\frac{d\omega}{2\pi}\,
  Y(\omega)(-i\omega) \frac{\phi(\omega)}{2e} e^{-i\omega t}+2e E_J\sin \phi(t)=0
\label{eq:saddle_point}
\end{equation}
and is equivalent to condition of current conservation.
The admittance $Y(\omega)$ here determines the time-dependent response of current on voltage $\dot{\phi}/2e$. 

%specify
We specify to the case of a single resonance mode,  such that the impedance near the resonant frequency $\Omega_0$ is given by Eq. \ref{eq:resonant-impedance}.
To achieve the conditions of the parametric resonance, we tune the d.c. bias voltage to $V_b=\hbar\Omega/e$ 
corresponding to the Josephson frequency $2\Omega$ close to the double of the resonant frequency  $\Omega_0$. 
The detuning $\nu_0 \equiv \Omega-\Omega_0$ is assumed to be much smaller than $\Omega_0$. To implement this assumption, we introduce a slow  
complex variable  $\varphi(t)$, an amplitude of the resonant field, and express the original variable $\phi$ as  
\begin{equation}
\label{eq:rotating-wave}
\phi^\pm(t) = 2 \Omega t + \mathop{\rm Re} [ e^{-i \Omega t}\varphi^\pm(t)].
\end{equation}
thereby disregarding its Fourier components far from $\pm\Omega$. This is equivalent to a rotating-wave approximation.

We substitute $\phi(t)$ to Eq. \ref{eq:actionj} in the form (\ref{eq:rotating-wave}) and average it over the period of resonant oscillations to obtain local-in-time action for the slow variable $\varphi(t)$,
\begin{align}\label{eq:slow_actionj}
  S_J = &\int dt \left(\bar{S}_J(\varphi^+(t)) - \bar{S}_J(\varphi^-(t))\right); \\
  \bar{S}_J(\varphi) = &\frac{E_J}{2} \frac{J_2(|\varphi|)}{|\varphi|^2}\left(\varphi^2 +(\varphi^*)^2\right)
\end{align}

We also express the environment part of action in terms of the slow variable,
\begin{equation}\label{eq:actionenv-slow} 
S_\text{env} = \frac{i}{8\pi G_Q}
  \int
  \frac{d\omega}{2\pi} \sum_{\alpha,\beta =\pm} (\phi_\omega^\alpha)^*
  M_{\alpha\beta}(\omega) \phi_\omega ^\beta
\end{equation}
with 
\begin{align}\label{eq:M}
  Z_0M(\nu) = -i(\nu+\nu_0 ) \begin{pmatrix}
    1&  0 \\
    0& -1  
  \end{pmatrix} +\\
 \Gamma \begin{pmatrix}
    n_\Omega +\tfrac{1}{2} &  -n_\Omega  \\
    - (n_\Omega +1) & n_\Omega +\tfrac{1}{2}  
  \end{pmatrix}  
\end{align}
where we introduce integration over "low" frequencies $\nu$. The above expression can be rewritten in local-in-time form, that contains local time derivatives of the fields only, 
\begin{align}
\label{eq:en-action-local}
  S_\text{env} = \frac{i }{16\pi G_Q Z_0}
  \int\!
 dt \, \left( \varphi^{+*} \partial_t \varphi^{+} -\varphi^{-*} \partial_t \varphi^{-}  \right. \nonumber\\
 -i \nu_0 \left(\varphi^{+*} \varphi^{+} -\varphi^{-*} \varphi^{-}  \right)\nonumber\\
 + \Gamma\left(n_\Omega +\tfrac{1}{2}\right) \left(\varphi^{+*} \varphi^{+} +\varphi^{-*} \varphi^{-} \right) \nonumber\\
 \left.
 - \Gamma \left(n_\Omega \varphi^{+*} \varphi^{-}+(n_\Omega+1)\varphi^{-*} \varphi^{+}\right) \right).
  %\label{eq:quadratic-action-environment}
\end{align}
This form of the (part of the ) action is proficient to establish a relation with traditional optical techniques.
If we rescale the variable $\phi$, $b = \phi /(4\sqrt{G_Q Z_0})$, the rescaled variables $b^*,b$ will provide path-integral representation of creation/annihilation operators $\hat{b}^\dagger, \hat{b}$ satisfying the standard commutation relations. Since the action is local in time containing the derivatives only, the path-integral can be solved with an evolution equation. In case under consideration, this evolution equation is the Bloch equation in the rotating-wave approximation for density matrix in $\hat{b}^\dagger, \hat{b}$ variables. It assumes the standard form implemented, for instance, in \cite{VyasSingh89}. We do not outline this equation here since we proceed with a different method.

Within this approximation, the "classical" equation \eqref{eq:saddle_point} that corresponds to the saddle-point solution of the action can be written as 
\begin{align}
\label{eq:classical_threshold}
\frac{d\varphi}{dt} = \left(i\nu_0 - \frac{\Gamma}{2}\right)\varphi 
+i (8\pi E_J G_Q Z_0) \nonumber\\
\times \left(\varphi^*\frac{2J_2(|\varphi|)}{|\varphi|^2} -\varphi ((\varphi^*)^2 +\varphi^2  ) \frac{J_3(|\varphi|)}{2|\varphi|^3}\right)
\end{align}
%\begin{equation}
 % \frac{\nu_0+i\Gamma/2 }{2\pi G_Q Z_0 E_J }+
 %\frac{J_1(|\varphi|)}{|\varphi|}\opcl{\frac{\varphi^*}{|\varphi|}}^3 - \frac{J_3(|\varphi|)}{|\varphi|}\frac{\varphi}{|\varphi|}=0.
%\label{eq:classical_threshold}
%\end{equation}
%
(See e.g. \cite{Likharev}). 
The Eq. \eqref{eq:classical_threshold} has stationary stable non-trivial  solutions $\varphi\ne0$, provided the Josephson energy exceeds a threshold $E_J \geq E_{\rm t}$, 
$E_{\rm t} = \Omega/4\pi G_Q|Z(\Omega)| = (\Gamma^2 +4\nu^2_0)^{1/2} /(8\pi G_Q Z_0)$. These solutions give coherent emission at the frequency $\Omega$. Below the threshold, quantum fluctuations enable emission of photon pairs resulting in incoherent radiation with linewidth $\simeq \Gamma$.

We will restrict our consideration to the situation below the threshold. It is essential to note that the typical quantum fluctuation of $\varphi$ remains small below the threshold, $(\delta \varphi(t))^2\ll 1$. This is guaranteed by the fact that the impedance $Z_0$ is small, $(\delta \varphi(t))^2 \simeq Z_0G_Q$. The fluctuations eventually grow at approaching the threshold. However, they will become of the order of $1$ only in a close vicinity of the transition estimated as $|E_J - E_t| \simeq (Z_0G_Q)E_t \ll E_t$. Therefore, almost everywhere below the threshold  we may expand $S_J$ in Taylor series in $\varphi$ keeping  the leading  quadratic term only, 
 \begin{equation}
S_J=\frac{E_J}{16}\int\! dt \left( (\varphi^+)^2 - (\varphi^-)^2+ c.c. \right)      
 \label{eq:quadratic-action-Josephson}
 \end{equation}
We conclude that below the threshold the total action is quadratic describing Gaussian fluctuations 
of the field. One could get an impression of rather trivial statistics. Indeed, if we  consider the statistics of the field itself, as it has been done in \cite{Kindermann} for a general linear electric circuit, we would end up with normal distributions. The point is that we are interested in the statistics of the photon flow, a variable that is quadratic in field. This leads to a non-trivial non-Gaussian statistics.

Our goal is thus to describe the full counting statistics of photons emitted from the resonator. 
Most general characteristic function of these statistics is expressed as \cite{VyasSingh89,ShortReview,QuantumTransport}
\begin{align}
\label{eq:gencharfunct}
{\cal Z}(\{\chi(t)\}) = {\rm Tr} \left[ {\rm Texp}\left(-i \int dt \hat{I}(t) \frac{\chi(t)}{2}\right) \hat{\rho}_{-\infty} \right. \nonumber \\
\left. {\rm \bar{T}exp}\left(-i \int dt \hat{I}(t) \frac{\chi(t)}{2}\right)\right]
\end{align}
where $T(\bar{T})$ denotes (anti)time ordering of the exponents, $\hat{I} \equiv \partial_t \hat{N}$ is the operator of photon flow from the resonator, $\hat{N}$ being the photon number operator, $\hat{\rho}_{-\infty}$ being the density matrix . Indeed, expansion of  (\ref{eq:gencharfunct}) in powers of $\chi(t)$ delivers the time-dependent correlators of the operators $\hat{I}$.  The characteristic function can be presented by a path integral over the field configurations with the Keldysh action modified by the counting field $\chi(t)$. 
%This modification in fact amounts to an elongation of the time derivative in Eq. \ref{eq:en-action-local},
%\begin{align}
%\varphi^{*+} \partial_t \varhi^+ \mapsto \varphi^{*+} \left (\partial_t + \frac{\dot{\chi}}{2} \right)\varhi^+ \nonumber \\
%\varphi^{*-} \partial_t \varhi^- \mapsto \varphi^{*+} \left (\partial_t + \frac{\dot{\chi}}{2} \right)\varhi^-
%\end{align}
%Indeed, the expansion in $\dot{\chi}$ in this case delivers the correlators of $\hat{N}$. It is instructive at this point to perform a  contour-dependent unitary transformation of the fields $\varphi^{\pm}(t) \to \exp(-i \mp \chi(t)) \varphi^{\pm}(t)$ \cite{Transformation} that cancels the derivative elongation while modifying the cross terms like $\varphi^{*+}\varphi^{-}$. 
(see \cite{Levitov, ShortReview} for fermion case, \cite{KindermannBeenakkerNazarov} for photon case).
 With this, the only modified term in the action is the third one in (\ref{eq:M}) , and the modification reads 
\begin{align}
\Gamma \begin{pmatrix}
    n_\Omega +\tfrac{1}{2} &  -n_\Omega  \\
    - (n_\Omega +1) & n_\Omega +\tfrac{1}{2}  
  \end{pmatrix}  \mapsto \nonumber \\\Gamma \begin{pmatrix}
    n_\Omega +\tfrac{1}{2} &  -n_\Omega e^{-i\chi(t)} \\
    - (n_\Omega +1)e^{i\chi(t)} & n_\Omega +\tfrac{1}{2}  
  \end{pmatrix}  
\label{eq:modification}
\end{align}
This form of the modification is suggestive and can be derived heuristically. The fact that counting field enters the action in the form of exponents guarantees the integer number of counts. If one rewrites the action in the form of master/Bloch equation for an extended density matrix \cite{QuantumTransport,Bagrets,Romito}, the modification concerns the terms that describe transitions with emission ($\Gamma \exp(i\chi) (n_\Omega +1)$) or absorption ($\Gamma \exp(i\chi) n_\Omega $) of a single photon, filling factor of the environment photons entering the rates of these transitions in an expected way.

The time-dependent counting field in the action is a parameter, that can be chosen at will. A common choice is a piecewise-constant $\chi(t)$, $\chi(t) =\chi$ with a time interval $(0, \tau)$. Computed ${\cal Z}(\chi)$ becomes in this case the characteristic function of the probability distribution of emitting $N$ photons
 within this time interval ,
\begin{equation}
 P(N) = \int\!\frac{d\chi}{2\pi}\; {\cal Z}(\chi)\; e^{-i\chi N}
\label{eq:emission_probability}
\end{equation}
and the cumulants  of $N$ are obtained via the differential relation
\begin{equation}
\langle \avg{N^m} \rangle = \partial_{i\chi}^m \ln({\cal Z}(\chi)) |_{\chi=0}.
\label{eq:moments}
\end{equation}
In this work, we will concentrate on the low-frequency limit of the FCS assuming $\tau$ to be much bigger than the typical waiting time of the phonon emission and disregarding the contribution associated with the ends of the interval that does not depend on $\tau$. With this,
\begin{equation}
\ln({\cal Z}(\chi)) = - \lambda(\chi) \frac{\Gamma}{2} \tau
\end{equation}
all information about the statistics being incorporated into a dimensionless function $\lambda(\chi)$. The advantage of this assumption is that one can disregard the time-dependence of $\chi(t)$ in the action that automates the evaluation of the path integral.

So far we have assumed an ideal efficiency of counting. If the statistics in this limit are known, one can easily obtain the results for any efficiency $f$. The method is to replace in all expressions for characteristic functions 
\begin{equation}
\label{eq:map-efficiency}
\exp(i\chi) \mapsto 1 +f(\exp(i\chi)-1).
\end{equation}
It is simple to justify this heuristically. One can split the whole damping rate $\Gamma$ into undetectable losses $\Gamma_1$ and losses detected, $\Gamma_2$, $f = \Gamma_2/(\Gamma_1+\Gamma_2)$. $\Gamma_1$ and $\Gamma_2$ both provide independent additive contributions to the action, and only the second one is modified with the counting field.

\section{Full counting Statistics}
\label{sec:statistics}

To represent the resulting action in a compact form, we introduce four independent scalar real fields corresponding to the complex fields $\varphi$ for positive and negative $\nu$ at forward/backward part of the contour. We group those in a 4-vector $\psi_{\nu}= \left[\varphi^+_{\nu},\varphi^-_{\nu},\opcl{\varphi^+_{-\nu}}^*,\opcl{\varphi^-_{-\nu}}^*\right]^T$, such that the modified action can be expressed compactly as a $4\times 4$ quadratic form in $\psi_{\nu}$ and $\psi_{\nu}^*$,
\begin{align}
S=&\ \frac{i}{16\pi G_Q Z_0}\int\frac{d\nu}{2\pi}\opcl{\psi_{\nu}^{\alpha}}^*A_{\nu}^{\alpha\beta}\psi_{\nu}^{\beta};\\
A_{\nu}  =&\left( \begin{array}{cc} M(\nu,\chi) & \Delta \cr \Delta & M^T(-\nu,\chi)\end{array}\right)
\end{align}
where $2\times2$ matrix $M(\nu,\chi)$ is given by Eq. \ref{eq:M} with the modification (\ref{eq:modification}),
and 
\begin{equation}
\Delta = i \frac{E\Gamma}{2} \left( \begin{array}{cc} 1 & 0 \cr 0 & 1\end{array}\right)
\end{equation}
where we have introduced a convenient dimensionless measure of Josephson energy $E=\ 8\pi G_QZ_0 (E_J/\Gamma)$.

We take  the path integral. Since it is Gaussian, the computation amounts to evaluation of the determinant of the quadratic form. Since $\chi$ is assumed to be time-independent, the quadratic form separates for each frequency. It is convenient  to introduce discrete frequencies spaced with $2\pi/\tau$. Then, the integrals over the fields at each discrete frequency are Gaussian integrals proportional to the inverse of the determinant of matrix $A_{\nu}$. We transform the resulting product of determinants into the exponent of a sum and go to the continuous limit in this sum recovering the integral over the frequencies. The result in the integral form reads
\begin{equation}
{\cal Z}(\chi)= \exp\opcl{-\tau\int^{\infty}_{0}\frac{d\nu}{2\pi}\ \ln\opcl{\frac{\det\opcl{A_{\nu}(\chi)}}{\det\opcl{A_{\nu}(\chi=0)}}}}
\label{eq:partition}
\end{equation}

It is convenient to introduce dimensionless variables: $\ti{\nu}=2\nu/\Gamma$, $\ti{\nu}_0=2\nu_0/\Gamma$,  and a dimensionless parameter $d=1+\ti{\nu}_0^2-E^2$. The latter is important and enters most results presented below. The parameter $d$ is positive, $d=0$ at the instability threshold, $d = 1+\ti{\nu}_0^2 >1$ at $E_J=0$, this is, in the absence of the parametric driving. With this, the statistics are expressed in a simple integral form, 
\begin{align}
\label{eq:counting}
\lambda(\chi)=&\ \int^{\infty}_{0}\frac{d\ti{\nu}}{2\pi}\ \ln\opcl{1+\frac{4z(\chi)}{p(\ti{\nu})}}\\
z(\chi)=&\ E^2 n_{\Omega}^2\opcl{1-e^{-2i\chi}}+\\
\ &\ E^2 (1+n_{\Omega})^2\opcl{1-e^{2i\chi}}\notag\\
p(\ti{\nu})=&\ \ti{\nu}^4+\ti{\nu}^2\ 2\opcl{2-d}+d^2
\end{align}
We take the integral over the frequencies to arrive at
\begin{equation}
\label{eq:lambda0}
\lambda(\chi)=-1+\sqrt{1-\tfrac{d}{2}+\sqrt{\opcl{\tfrac{d}{2}}^2+z(\chi)}}.
\end{equation}
This gives the FCS at arbitrary temperatures. The structure of $z(\chi)$ suggest that photons are emitted/absorbed in pairs ($\exp(\pm i2\chi)$ factors). Each emission/absorption probability is affected with the filling factors as expected: absorption probability is proportional to $n^2_\Omega$ (two photons), while emission is stimulated  with a factor $(1+n_\Omega)^2)$. In the limit of small $E \ll d$ one can expand $\lambda$ in terms of $z$ to arrive at
\begin{equation}
\lambda(\chi) = - \frac{E^2}{2d} n^2_\Omega (e^{-2i\chi} -1) - \frac{E^2}{2d} (1+n_\Omega)^2 (e^{2i\chi} -1)
\end{equation}  
This limit corresponds to the independent pair emission/absorption acts, that can be thus regarded as uncorrelated events. This gives Poissonian distribution of pair counts. The rate of pair emission(absorption) is $\Gamma_e = \Gamma (E^2/4d) (1+n_\Omega)^2)$($\Gamma_a = \Gamma (E^2/4d) n_\Omega^2$) and is much smaller than $\Gamma$ under assumptions made. Upon increasing $E$, the correlations between the pair events set in. No event would take place at zero $E$.

It may seem strange that at finite temperature no single-photon events are manifested in the FCS we present. Such events do take place, even in the absence of the parametric drive $E$: photons from the environment are randomly absorbed/emitted by/from the resonator. The point is that we concentrate here on the statistics in zero-frequency limit, and count all photons emitted/absorbed. In terms of cumulants of counts within a finite time interval $\tau$, we thus concentrate on the part of a cumulant that grows $\propto \tau$. Such parts are absent for the single-photon statistics mentioned, and therefore these events do not contribute to the FCS we describe. An alternative way to understand this is to notice that without parametric drive the detector is in thermal equilibrium with the resonator. It is known that in this case it will not produce any (count) signal.  

From now on we will focus on the quantum limit, that is, on the case of vanishing temperature $k_BT\ll\hbar\Omega$, so that $n_{\Omega}\mapsto 0$.
 In this case, $z(\chi)= E^2 \opcl{1-e^{2i\chi}}$. This indicates that in this limit only pair emissions take place. The FCS expression reduces to

\begin{equation}
\label{eq:lambda}
\lambda(\chi)=-1+\sqrt{1-\tfrac{d}{2}+\sqrt{\opcl{\tfrac{d}{2}}^2+E^2\opcl{1-e^{2i\chi}}}}
\end{equation}

This is one of the main results of this paper. In the limit of zero detuning $d=1-E^2$ and $\tau \to \infty$ this coincides with the results of Ref. \onlinecite{VyasSingh89}. 

\subsection{Average intensity and Fano factor}
\label{sec:moments}

Let us evaluate  the first two moments of the statistics derived: 
the average intensity  $\bar{I} = \avg{N}/\tau$ and the intensity noise $S_I  = \avg{\avg{N^2}}/\tau$. Expanding Eq. \ref{eq:lambda} in $\chi$ gives 
\begin{align}
\label{eq:N_average}
\frac{\avg{N}}{\tau}= -\frac{\Gamma}{2}\left.\frac{\partial \lambda(\chi)}{\partial(i\chi)}\right\vert_{\chi=0}
                                  = \frac{\Gamma}{2}\frac{E^2}{d}
\end{align}
(illustrated in Fig. \ref{fig:setup}) and 
\begin{align}
\label{eq:variance}
\frac{\avg{\avg{N}}}{\tau}= -\left.\frac{\partial^2 \lambda(\chi)}{\partial(i\chi)^2}\right\vert_{\chi=0}
                                            = 2\frac{\avg{N}}{\tau} + \frac{\Gamma}{2}\; E^4\; \frac{\opcl{4+d}}{d^3}\ .
\end{align}
 From this we
can determine the Fano factor $F=\avg{\avg{N^2}}/\avg{N} = S_I/\bar{I}$. This number is significant in electron or photon counting statistics giving an estimate of a number of particles that correlate with each other. We obtain
\begin{equation}
\label{eq:Fano}
F=2+E^2\frac{4+d}{d^2}
\end{equation}
We see that Fano factor is 2 in the limit of weak parametric driving and diverges upon approaching the instability threshold $d\to 0$. It is instructive to re-write this expression in terms of average number of photons in the resonator $\bar{N} = \bar{I} /\Gamma = E^2/2d$ and detuning $\tilde{\nu}_0$. 

\begin{equation}
\label{eq:Fano-N}
F=2+2 \bar{N} \left(1+ \frac{4(2\bar{N} +1)}{1+\tilde{\nu}^2_0}\right)
\end{equation}

The increased Fano factor is surely due to photon bunching. A naive picture of such bunching would presume that $N$ photons (in the resonator) stimulate emission of one another, that is  $F \propto \bar{N}$. Hanbury-Brown-Twiss relation also supports such estimate. In the limit of large detuning $\tilde{\nu}_0 \to \infty$ we indeed recover $F=2(1+\bar{N})$. However, generally it is not so: near the instability threshold ($\bar{N} \to \infty$) $F = 16 \bar{N}^2 / (1+\tilde{\nu}^2_0) \simeq \bar{N}^2$. The number of photons correlated exceeds by far the number of photons present in the resonator! This is specific for the parametric resonance.  

If only a fraction of the emitted photons is measured by the detector, 
the correlation decreases.  With the aid of Eq. \ref{eq:map-efficiency} the
Fano factor  measured can be expressed in terms of 
the Fano factor at absolute efficiency,
\begin{equation}
\label{eq:partial_Fano}
F(f)=1+f(F-1)=(1+f)+E^2\frac{4+d}{d^2}f
\end{equation}
It approaches $1$ in the limit of small efficiency.

\begin{figure}
\centerline{\includegraphics[width=0.9\linewidth]{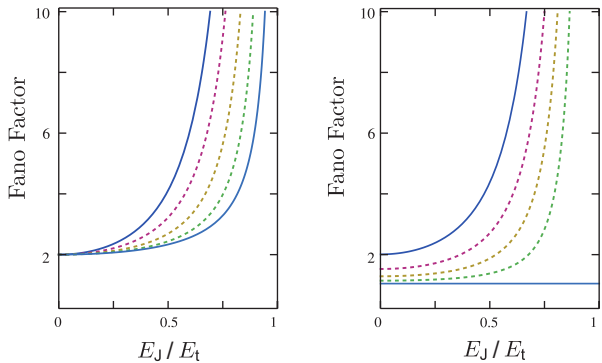}}
\caption{Fano factor versus Josephson energy $E_J$. The energy $E_{\rm t}$ corresponds to the instability threshold. Left panel: dependence on detuning $\tilde{\nu}_0$. The curves from  bottom to  top correspond to $\tilde{\nu}_0=0, 0.5,1,2,\infty$, solid curves corresponding to  the extreme values $\tilde{\nu}_0=0,\infty$ . Absolute efficiency $f=1$ is assumed.  Right panel: Dependence of $F$ on detection efficiency $f$ at $\tilde{\nu}_0=0$. The curves from top to bottom correspond to the efficiencies $f=1, 0.5, 0.25, 0.1, 0$, solid curves corresponding to  the extreme values $f=1,0$. }
\label{fig_Fano}
\end{figure}

The dependence of the Fano factor on the parametric drive is illustrated in Figure \ref{fig_Fano}.

\subsection{Big deviations}
\label{sec:big-deviations}

The FCS expression (\ref{eq:lambda}) can be employed to find with the exponential accuracy the probability of big deviations of $I$ from its expectation value $\bar{I}$ (see e.g. \cite{KindermannBeenakkerNazarov}).  To do this, one evaluates integral in Eq. \ref{eq:emission_probability} at $N=I \tau$ in the saddle-point approximation to obtain 
\begin{equation}
P(I) = \int\!\frac{d\chi}{2\pi} e^{-i\chi I\tau -\frac{\Gamma\tau}{2}\lambda(\chi)} \propto e^{-\frac{\Gamma \tau}{2} {\cal L}(I)}
\end{equation}
with 
\begin{equation}
{\cal L}(x) = \mathop{{\rm min}}_{\mu} \left(\frac{I}{\Gamma/2} \mu + \lambda(-i\mu)\right).
\end{equation}
For any FCS expression, ${\cal L}({\bar I})=0$, and achieves a minimum there. The quadratic expansion near the minimum corresponds to a Gaussian distribution  of small deviations from the expectation value. The probability of big deviations is not Gaussian although exponentially small.
 
\begin{figure}
\centerline{\includegraphics[width=0.95\linewidth]{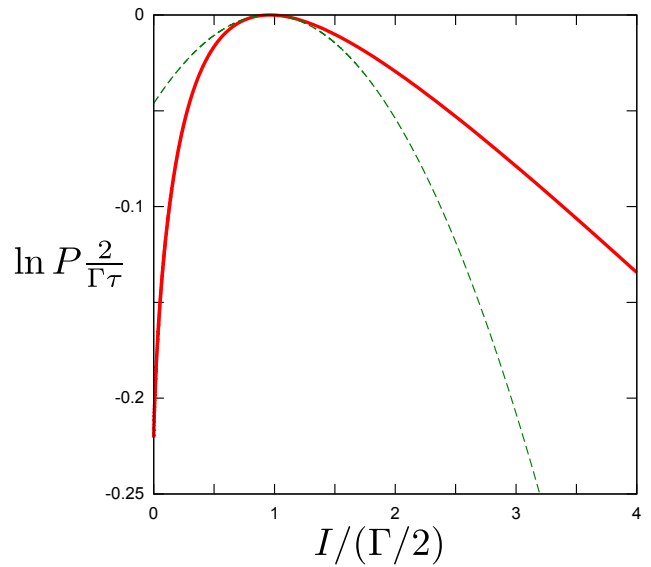}}
\caption{Probability of big deviations of $I$ for $E=0.7$, $\nu_0=0$. ($\bar{I} \approx 0.96 (\Gamma/2)$). Dashed line: Gaussian approximation valid for small deviations from the expectation value.}
\label{fig:prob}
\end{figure}

A typical dependence of $\ln P$ on $I$ is shown in Fig. \ref{fig:prob} along with its Gaussian approximation. The probability is lower than the Gaussian approximation at $I < \bar{I}$ and higher otherwise.
 
 A feature worth discussing is that  the probability to emit no photons ($I=0$) is finite and given by
\begin{equation}
- \frac{2}{\Gamma \tau} \ln P  = \lambda(i\infty) = -1+\sqrt{1-\tfrac{d}{2}+\sqrt{\opcl{\tfrac{d}{2}}^2+E^2}}
\end{equation} 
Rather counterintuitively, this probability remains finite even at approaching the threshold where $\bar{I} \to \infty$,
\begin{equation}
-\ln P \to \frac{\Gamma\tau}{2} \left(-1+\sqrt{1-\sqrt{1+\tilde{\nu}^2_0}}\right)
\end{equation}

\begin{figure}
\centerline{\includegraphics[width=0.95\linewidth]{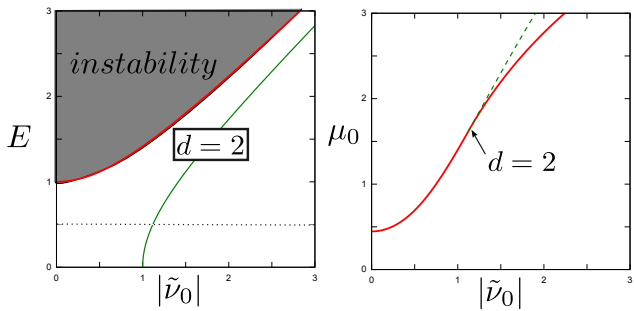}}
\caption{Probability of big deviations $I \gg \bar{I}$ (Eq.\ref{eq:plargeI}) exhibits a peculiarity at $d=2$. Left: The "transition" line $d=2$ in the plane of  parametric drive $E$ and detuning $\tilde{\nu}_0$ separates the plane into the regions of small and large detuning. Right: The coefficient $\mu_0$ plotted along the dotted line in the left pane. To make the peculiarity visible, dashed curve gives the analytical continuation from $d<2$ to $d>2$.}
\label{fig:hypandmu0}
\end{figure}

Another feature worth discussing is the probability at $I \gg \bar{I}$. The log of the probability appears to be proportional to $I$,
\begin{equation}
\label{eq:plargeI}
-\ln P = I \tau \mu_0
\end{equation}
where $-i\mu_0$ gives the position of the singularity of $\lambda(\chi)$ in the plane of complex $\chi$. The singularity comes either from the inner or outer square root in (\ref{eq:lambda}). Owing to this, $\mu_0$ exhibits a peculiarity (discontinuity of the second derivative) at $d=2$ where the square roots merge into a $1/4$ singularity (Fig. \ref{fig:hypandmu0})
\begin{equation}
\mu_0 =\left\lbrace \begin{array}{ll} \frac{1}{2} \ln\left(1+\frac{d^2}{4E^2}\right) & {\rm if } \ d<2 \cr
\frac{1}{2} \ln\left(1+\frac{d-1}{E^2}\right)  & {\rm if } \ d>2 \end{array}\right.
\end{equation}
The condition $d=2$ or, equivalently, $E = \sqrt{\tilde{\nu}_0^2 -1}$ gives thus a "transition" line that separates the parameter regions with large and zero detuning (Fig. \ref{fig:hypandmu0}).

It is also possible to find the next-to-the-leading term in the asymptotic expression (\ref{eq:plargeI}), an offset of linear asymptotics visible in Fig. \ref{fig:hypandmu0},  so the asymptotics become
\begin{align}
-\ln P = I \tau \mu_0 - C; \nonumber \\
C = \left\lbrace \begin{array}{ll} 1 - \sqrt(1-d/2) & {\rm if } \ d<2 \cr
1 & {\rm if } \ d>2 \end{array}\right.
\end{align}

\section{interpretation: bursts}
\label{sec:interpretation}

\begin{figure}
\centerline{\includegraphics[width=0.9\linewidth]{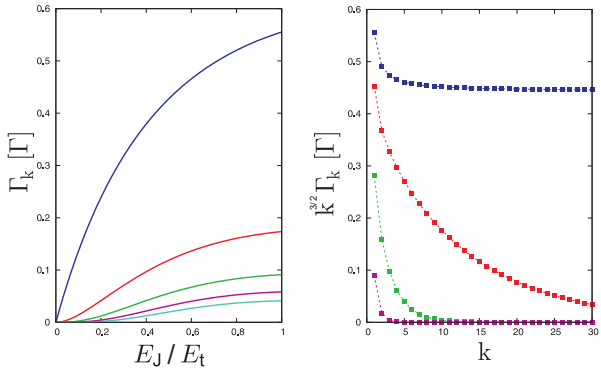}}
\caption{Rates of the bursts.  Right panel: The curves from top to bottom give the rates $\Gamma_1$ through $\Gamma_5$ at $\nu_0=0$ versus $E$, $E=1$ is the threshold . The rates are plotted as a function of Josephson energy scaled with respect to the energy $E_{\rm t}$, corresponding to the instability threshold. Left panel: The dependence of $\Gamma_k$ (normalized on the power-law $k^{-3/2}$) on the number of photon pairs in a burst. The curves from top to bottom correspond to Josephson energies $E_J/E_t=1,0.75,0.5,0.25,0.1$. Integer values of $k$ are marked by squares. }
\label{fig:rates}
\end{figure}

To understand better the FCS (\ref{eq:lambda}), let us give an interpretation of these statistics. Let us note that 
the integral form of $\lambda(\chi)$, Eq. \eqref{eq:counting} permits 
an expansion in powers of $e^{2i\chi}$, 
\begin{align}
\label{eq:bursts}
\lambda(\chi)=&\ \int^{\infty}_{0}\!\frac{d\ti{\nu}}{2\pi}\; 
						\ln\opcl{1+\frac{4E^2}{p(\ti{\nu})}}-\\
\ &\ \displaystyle\sum^{\infty}_{k=1}\frac{e^{2ik\chi}}{k}\; 
			\int^{\infty}_{0}\!\frac{d\ti{\nu}}{2\pi}\; 
							\opcl{\frac{4E^2}{p(\ti{\nu})+4E^2}}^k\notag
\end{align}
We rewrite it in the form 
\begin{equation}
-\frac{\Gamma}{2}\lambda(\chi) = \sum_{k=1}^\infty \Gamma_k \left(\exp(i2k\chi)-1\right).
\end{equation}
This suggests that photons are emitted in course of uncorrelated events, bursts, each accompanying $k$ photon pairs. 
The rate of a $k$-burst is given by
\begin{equation}
\label{eq:burst_rates}
\Gamma_k=\frac{\Gamma}{2k}\int^{\infty}_{0}\frac{d\ti{\nu}}{2\pi}\ \opcl{\frac{4E^2}{p(\ti{\nu})+4E^2}}^k \end{equation}
Analytical expressions for $\Gamma_k$ become increasingly complicated with increasing $k$ and we do not give them here.
 Their dependence on $E$ is illustrated in Fig. \ref{fig:rates}. At small $E$, $\Gamma_k \simeq E^{2k}$ as expected for the rate of an event encompassing $2k$ photons. 
 
 Note that the rates do not diverge at the threshold: rather, they saturate at finite value that decreases with increasing $k$.
To reconcile this with divergence of the radiation intensity at the threshold, let us determine the asymptotic behavior of $\Gamma_k$ in the limit of large  $k$.  The integral in Eq. \ref{eq:burst_rates} is contributed by minima of $p(\nu)$, and can be approximated by a Gaussian integral. There is a single minimum at $\tilde{\nu}=0$ if $d<2$ and two minima at $\tilde{\nu} = \pm\sqrt{d-2}$. The integration gives  
\begin{equation}
\label{eq:asymptotic_rates}
%\Gamma_{k\gg 1}\sim \frac{\Gamma}{8\sqrt{\pi}}\ \frac{1}{k^{3/2}}\ \opcl{\frac{1}{1+\opcl{\frac{d}{2E}}^2}}^k \sqrt{\frac{d^2+4E^2}{ 2\opcl{2-d}} }
\Gamma_{k\gg 1}\sim \frac{\Gamma}{8\sqrt{\pi}}\ \frac{1}{k^{3/2}}\ \exp\left(-2\mu_0 k\right)  \xi_0,
\end{equation}
where 
\begin{equation}
\xi_0 = \left\lbrace\begin{array}{ll} 
\sqrt{\frac{d^2+4E^2}{ 2\opcl{2-d}} } & {\rm if} \ d<2 \cr
2 \sqrt{\frac{d-1+E^2}{d-2}} & {\rm if} \ d>2 \end{array}\right.
\end{equation}
Comparing this with the probability of big deviations, we conclude that the big deviation is most likely a result of a single burst encompassing $k = I \tau$ photons during the observation interval.

Near the threshold, these asymptotics read
\begin{equation}
\Gamma_{k\gg 1} \sim E_t \frac{\Gamma}{8\sqrt{\pi}}\ \frac{\exp\left(-k\frac{E}{\bar{N}^2}\right)}{k^{3/2}} .
\end{equation}

We see that at the threshold the relative probabilities of $k$-bursts satisfy power law $k^{-3/2}$. Although the probability of big bursts is low, their contribution to the radiation intensity  is high such that the average intensity diverges. Below the threshold, the power-law distribution is cut at $k \simeq \bar{N}^2$. This gives an estimate of the typical burst size contributing to the intensity, which is in agreement with an earlier estimation obtained from the Fano factor (Eq. \ref{eq:Fano}).

The $k$-dependence of the rates at not-so-big $k$ is illustrated in Fig. \ref{fig:rates}. We see that the asymptotics is reached at rather low $k$.

\section{Limits}
\label{sec:limit}
We consider here two specific cases of the FCS under consideration: the limits of large detuning $\tilde{\nu_0} \gg 1$ (large $d$), and small $d$ (the vicinity of the threshold).
\subsection{Large detuning}
In this case, one can assume $d \gg 1$ almost everywhere in the sub-threshold region except a close vicinity $(E_t - E)/E_t \simeq \tilde{\nu}^{-1}_0$  of the threshold, and $E \lesssim \sqrt{d}$.
In this case, the inner square root in Eq. \ref{eq:lambda} can be expanded in 
$E^2$. The resulting FCS depends only on $\bar{N}$ and reads
\begin{equation}
\label{eq:lamda-large-detunings}
\lambda(\chi) = -1 +\sqrt{ 1 + \bar{N}(e^{2i \chi}-1)}
\end{equation}
It corresponds to the 'naive' estimation of the Fano factor  $F=2(\bar{N}+1)$.
This form is very similar to FCS of incoherent light with a Lorentz-shaped spectral intensity \cite{KindermannBeenakkerNazarov,oldlight} with $\bar{N}$ replaced by the maximum filling factor of the photons in the light. The difference is that in our case the photons come in pairs rather than one-by-one ($\exp(i\chi) \to \exp(i2\chi)$).

The origin of this similarity is understood if we consider the spectral intensity of the pairs emitted. This is given by inverse of $p(\nu)$ and in the limit of large detuning consists of two narrow Lorentz-shaped lines centered at $\tilde{\nu} = \pm \sqrt{d}$.
In course of pair emission, each constitute of the pair appear in a separate line.
Since the lines do not overlap, the photon bunching takes place separately within each line and has the same form as in the single-photon case. 

\subsection{Vicinity of the threshold} 
\label{sec:vicinity}
The distance to the threshold is parametrized by $d\ll 1$, $d=0$ precisely at the threshold. We need to expand $\lambda(\chi)$ in $d$. The formal expansion, however, does not work resulting in expressions that are singular at $\chi=0$ and therefore cannot be associated with any probability distribution. To preserve analyticity in $\chi$, we need to explicitly address small $\chi \ll 1$.  To this end, we may expand  $\exp(2i\chi)$ in $\chi$ up to the first order. This disregards the discreteness of the photon flow, which is a valid approximation at time-scales exceeding $\bar{I}$. We rearrange terms to arrive at  
\begin{equation}
\label{eq:limit}
\lambda(\chi)= -1+\sqrt{1-\tfrac{d}{2}\opcl{1-\sqrt{1-8i\chi\;E^2/ d^2}}}
\end{equation}
From this it is clear that small $\chi \sim d^2/4E^2$ eventually determine the statistics. Now we can expanding in  $d$ to recover a simpler FCS expression
\begin{equation}
\label{eq:K_distribution}
\lambda(\chi) = \frac{d}{4}\opcl{-1+\sqrt{1-8i\chi\;E^2/d^2}}
\end{equation}

 To find the probability  $P(I)$  of big deviations, we take the integral in saddle point approximation
\begin{align}
P(N)\sim &\ \exp\left[-\frac{\tau\Gamma d}{16}\opcl{\frac{I}{\bar{I}}+\frac{\bar{I}}{I}-2}\right]
\end{align}
This  form has been discussed previously in the context of photon counting statistics\cite{KindermannBeenakkerNazarov}.
 The expression for the probability is valid only if exponentially small, that is, for the observation times $\tau\gg (\Gamma d)^{-1}$). This suggests the relevance of a long time-scale $\simeq  (\Gamma d)^{-1} \gg \Gamma^{-1}$ in the vicinity of the threshold.

\section{Time-dependent correlations}
\label{sec:time}
The burst interpretation outlined above would have been fine if the uncorrelated bursts could be regarded as instant events. In fact, the events are not instant: it takes time to emit $k$ pairs composing a burst. If $\bar{N} \simeq 1$ and $k \simeq 1$, this time is of the order of $\Gamma^{-1}$. Close to threshold , $\bar{N} \gg 1$ and the typical waiting time between pair emissions is short, $\simeq \bar{I}^{-1} \simeq \Gamma^{-1}$. However, a typical burst in this case encompasses $\bar{N}^2$ photons. This implies that the time required for a burst is actually long, $\simeq \Gamma^{-1} \bar{N}$, in agreement with the final remark in Subsection \ref{sec:vicinity}. The bursts are thus overlapped in time. Moreover, even in the limit of $\bar{N} \ll 1$ when the events are pair emissions that are well-separated in time, the constituents of the pair do not have to be emitted simultaneously. The low-frequency FCS computed does not provide direct information about such time correlations. 

Here, we will investigate the time correlations restricting to a simple case where we can proceed perturbatively. Let us choose the time-dependent counting field in the form 
\begin{equation}
\chi(t) =\chi_1 \Theta(t_1,t_1+dt_1) +\chi_ 2\Theta(t_2,t_2+dt_2),
\end{equation} 
where $\Theta(t_a,t_b) \equiv \Theta(t_b -t)\Theta(t-t_a)$. The counting field is thus piece-wise constant that is non-zero in two time intervals. If the duration of these time intervals is small, $dt_1,dt_2 \ll \bar{I}^{-1}$, the chance to have photon emissions within these intervals is small and can be computed perturbatively. In this case, the unperturbed action corresponds to $\chi=0$ and we expand in terms of the perturbation
\begin{equation}
S_{\text{int}} = -\frac{i\Gamma}{16 G_Q Z_0}  \int dt (\exp(i\chi(t))-1) \varphi^{+*}(t) \varphi^{-}(t)
\end{equation}
(see Eqs. \ref{eq:en-action-local}, \ref{eq:modification} ) We assume that the time distance between the intervals $\tau \equiv t_2 -t_1 \gg dt_{1,2}$ is much bigger than the interval durations.

Expansion of the cumulant-generating function $\ln {\cal Z}(\chi)$ up to the second order gives
\begin{align}
\ln {\cal Z}(\chi)  = &C_1 (\exp(i\chi_1)-1) d t_1 + C_2 (\exp(i\chi_1)-1) d t_1 + \nonumber \\
\ &C_{11}(\exp(i\chi_1)-1) ^2 (d t_1)^2 + \nonumber\\
\ &C_{22}(\exp(i\chi_2)-1) ^2 (d t_2)^2+ \nonumber \\
\ & C_{12}(\tau) (\exp(i\chi_1)-1)(\exp(i\chi_2)-1)
\end{align}
It is clear that $C_{1,2}$ give a chance of photon emission in the intervals, so that $C_{1,2} =\bar{I}$.
$C_{12}$ is of interest for us since it gives correlations between the emissions separated by time $\tau$: if an emission has occurred within the time interval $(t_1,t_1+dt_1)$, this increases a chance of emission within $(t_2,t_2+dt_2)$.  We express this increased chance in terms of a time-dependent excess intensity $I_{\text{ex}}(\tau)$,
$C_{12}(\tau) = \bar{I} I_{\text{ex}}(\tau)$.

From the other hand, the perturbations give
\begin{equation}
C_{12} = \ \frac{  \Gamma^{2}}{2^{10} (G_Q Z_0)^2}\langle\langle \phi^+(\tau)
        \phi^- (\tau) ^* \phi^+(0)
     \phi^- (0) ^* \rangle\rangle 
\end{equation}
Since the fluctuating field $\varphi$ is Gaussian, all correlators can be readily expressed in terms of the field propagator,
\begin{equation}
\label{eq:propagator-definition}
G_{\alpha\beta} (t,t') = \langle \psi_\alpha^*(t) \psi_\beta(t')\rangle ,
\end{equation}
that depends on time difference only, $G_{\alpha\beta}(t,t') = G_{\alpha\beta}(t-t')$. The quantity of interest is expressed as
\begin{align}
C_{12} = &\frac{  \Gamma^{2}}{2^{10} (G_Q Z_0)^2} 
\left(\langle \phi^+(t) \phi^- (0) ^*
     \rangle \langle \phi^- (t) ^* \phi^+(0) \rangle + \right.\notag  \\
    & \left.\langle \phi^+(t) \phi^+ (0)
          \rangle \langle \phi^- (t) ^* \phi^-(0)^* \rangle \right) =\notag\\
   = &\frac{  \Gamma^{2}}{2^{10} (G_Q Z_0)^2} \left( G_{34}(t) G_{21}(t) + G_{31}(t) G_{24}(t)\right)   
      \notag \\=& \frac{  \Gamma^{2}}{16} \left(  A_1 + A_2\right) 
\end{align}

The evaluation of the propagator is straightforward but cumbersome, so we present the details in the Appendix.

We calculate the correlation function of photon emission events separated by a time $t$.
%%
%\begin{align}\label{eq:cross}
 % \langle \langle N(t) N(0) \rangle \rangle 
  % = &\ \frac{  \Gamma^{2}\tau_1 \tau_2}{16g_0^2}\langle\langle \phi^+(t)
    %    \phi^- (t) ^* \phi^+(0)
     %\phi^- (0) ^* \rangle\rangle \nonumber\\
     %= &\ \frac{\Gamma^2 \tau_1 \tau_2}{16g_0^2} \langle \phi^+(t) \phi^- (0) ^*
     %\rangle \langle \phi^- (t) ^* \phi^+(0) \rangle + \notag  \\
%\ &\   \frac{\Gamma^2 \tau_1 \tau_2}{16g_0^2} \langle \phi^+(t) \phi^+ (0)
  %        \rangle \langle \phi^- (t) ^* \phi^-(0)^* \rangle \notag\\
     % =&\  A_1 + A_2 
%\end{align}
%
Two contributions to the correlator read ($\gamma_{\pm} =1\pm\sqrt{1-d}$)
\begin{align}
A_1=&\  \frac{E^4}{d^2\opcl{1-d}} \opcl{\gamma_- e^{-\gamma_+  \Gamma|\tau|/2} - \gamma_+ e^{-\gamma_-
   \Gamma|\tau|/2} }^2\notag\\
A_2=&\  \frac{E^4}{d^2\opcl{1-d}}\left[\opcl{\sum_{\pm}\gamma_\pm e^{-\gamma_\pm  \Gamma|\tau|/2} }^2- \frac{4\tilde\nu_0^2 d}{E^2}
              e^{-\Gamma |\tau|}    \right]\notag
\end{align}
The resulting excess intensity is therefore expressed as
\begin{align}
I_{\text{ex}}
 = \frac{\Gamma}{4}  \frac{E^2 d}{\opcl{1-d}}
 \left[ 
 \sum_{\pm} 
 \frac{e^{-\gamma_\pm \Gamma |t|}}{\left(\gamma_\pm \right)^2}
      - \frac{2 \tilde \nu_0^2 e^{-\Gamma|t|}}
      {E^2 d}\right]
      \end{align}

It is instructive to introduce the number of excess photons $n_\text{ex}(\tau)$ emitted within  the time interval $-|\tau|,|\tau|$  and obtained by the integration of the excess intensity $I_{ex}(\tau)$ over the time, $n_\text{ex}(0)=0$,
\begin{align}
n_{\text{ex}} = n_\infty -  \frac{E^2 d}{2\opcl{1-d}}\left[ 
 \sum_{\pm} 
 \frac{e^{-\gamma_\pm \Gamma |\tau|}}{\gamma_\pm }
      - \frac{2 \tilde \nu_0^2 e^{-\Gamma|\tau|}}
      {E^2 d}\right]
\end{align}

where the total number of excess photons $n_\infty \equiv n_{\text{ex}}(\infty)$ is related to the Fano factor (Eq. \ref{eq:Fano})
\begin{equation}\label{eq:condn}
  n_\infty = F -1.
\end{equation}
In the limit of small parametric drive $E \to 0$, $n_\infty =1$. This implies that each photon correlates with strictly one extra photon forming a pair. The time-dependence of the correlations in this limit is given by
\begin{equation}
n_{\text{ex}}(\tau) = 1- \exp(-\Gamma \tau)
\end{equation}
not depending on the detuning.

In the vicinity of the threshold, the correlations are big and mainly build up at the slow time scale $\simeq (\Gamma d)^{-1}$, 
\begin{equation}
n_{\text{ex}}(\tau) = F \left(1-\exp\left(-\frac{\Gamma d \tau}{2}\right)\right)
\end{equation}

\begin{figure}[h]
\centerline{\includegraphics[width=0.9\linewidth]{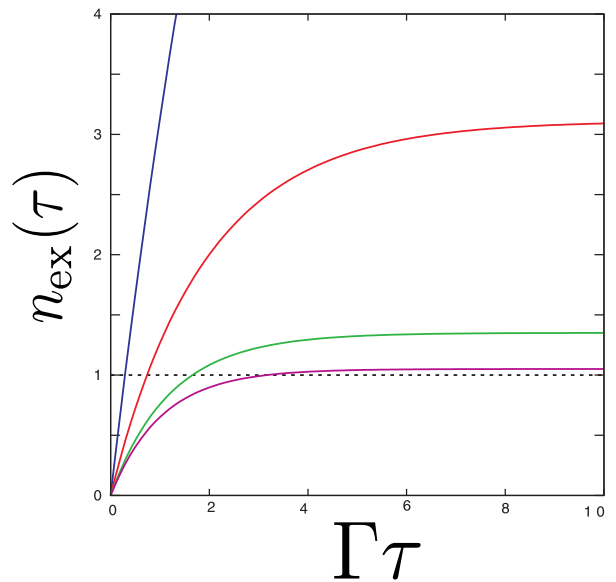}}
\caption{Time correlations of emissions: number of excess photons emitted in the time interval $(-\tau,\tau)$ provided the emission of a photon took place at $t=0$. From uppermost to lowermost, the curves correspond to Josephson energies $E=0.75,0.5,0.25,0.1$ at zero detuning.  The dotted line marks the value $n_{\text{ex}}=1$, 
one excess photon. The curve at small $E$ only exceeds $1$ only slightly, manifesting the fact that emissions occur in pairs. Emissions of pair constituents are separated by time interval $\simeq \Gamma^{-1}$. } 
\label{fig_time_correlation}
\end{figure}

The time-dependence of $n_{\text{ex}}(\tau)$ is illustrated in Fig. \ref{fig_time_correlation}.

\section{Frequency-resolved correlations}
\label{sec:resolved}
Experiments on FCS in Josephson parametric amplifier are plausible but, as all experiments on FCS, are difficult and long, requiring long times of data accumulation and careful characterization of extrinsic noises in measurement setups. The first experiments would most likely concern intensity noise, the second cumulant of FCS. However even in this case the measurement may be difficult since the measured signal has to be amplified and the amplifier brings in a substantial extra noise.  A common way to avoid such difficulties in the context of low-temperature measurement \cite{Glattli}
is to split a noisy signal into two parts and amplify them by independent amplifiers. The cross-correlation of two outputs will not be affected by the amplifier noise.

In our setup, it is convenient to split the signal in frequency domain. 
We introduce two detectors absorbing emitted photons with frequency-dependent efficiencies $f_{1,2}(\nu)$. This results in two intensity signals 
\begin{equation}
I_{1,2} = \int \frac{d\nu}{2\pi} f_{1,2}(\nu) \left(\frac{d I}{d\nu} \right),
\end{equation}
$dI/d\nu$ being intensity per frequency interval. For our setup, the average intensity per frequency interval reads (see Eq. (\ref{eq:counting}) )
\begin{equation}
\frac{d\bar{I}}{d\nu} = \frac{4 E^2}{4 \tilde{\nu}^2 + (\tilde{\nu}^2 - d)^2}.
\end{equation}
To describe the FCS of the two signals, we introduce two counting fields $\chi_{1,2}$.
With this, the $\chi$-dependent part of the action reads
\begin{align}
S = &\frac{-i \Gamma}{16 \pi G_Q Z_0}\int \frac{d \nu}{2\pi}  \left((e^{i\chi_1}-1)f_1(\nu) + \right. \notag\\
+&(e^{i\chi_2}-1)f_2(\nu) )(\varphi^{-}_{-\nu})^* \varphi^{+}_{-\nu}
\end{align}
We need only the cross-correlation of the intensities generally defined as
\begin{equation}
S_{12} = -\frac{\partial {\ln {\cal Z}}}{\partial \chi_1 \partial \chi_2} \tau^{-1}
\end{equation}
at $\chi_{1,2} \to 0$. Employing perturbations in $\chi_{1,2}$ we find 
\begin{align}
S_{12}= \int \frac{d \nu_1 }{2\pi}  \frac{d \nu_2 }{2\pi}   f_1(\nu_1) f(\nu_2) S(\nu_1,\nu_2)
\end{align}
where the intensity correlator is expressed in terms of the field averages  as
\begin{equation}
S(\nu_1,\nu_2) = \frac{\Gamma^2 \tau^{-1}}{2^8 (\pi G_Q Z_0)^2} \langle\langle (\varphi^-_{-\nu_1})^* \varphi^+_{-\nu_1} (\varphi^-_{-\nu_2})^*
  \varphi^+_{-\nu_2} \rangle\rangle
\end{equation}
 Expressing the correlator in terms of the field propagator, we find
 \begin{align}
S(\nu_1,\nu_2) = \frac{2 \Gamma^{-2}}{[4 \tilde{\nu}_1^2 + (\tilde{\nu}_1^2 - d)^2]^2} \notag \\
  \left[ E^2 (1+  E^2 + \tilde{\nu}_1^2 -2 \tilde \nu_0 -\tilde
  \nu_0^2)^2 \delta(\nu_1 + \nu_2) \right. \notag \\
  + 4  E^2 \left.\delta(\nu_1 - \nu_2) \right]
  \end{align}
  
 This defines the general form of the spectral-resolved correlations below the instability threshold. The correlations are delta-functional and
persist only for exactly equal or exactly opposite frequencies.  This seems to naturally describe bunching of the photons in the same frequency mode as well as emission of pairs with frequencies opposite owing to energy conservation. However, delta-functional correlations are an artifact of Gaussian approximation: taking non-linearities into account  would result in a smooth frequency dependence. Since we integrate over relatively wide frequency windows, the exact shape of the smoothed delta-functional peaks is not important for us.
 
\begin{figure}
\centerline{\includegraphics[width=0.9\linewidth]{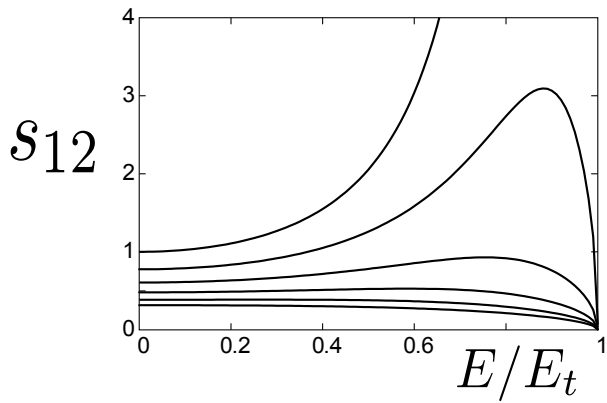}}
\caption{Normalized cross-correlation of intensities in two frequency windows (Eq.\ref{eq:two-wind}) versus $E$ ($\nu_0 = 0$). The separating frequency $\omega_s$ takes values $0, 0.1, 0.2, 0.3, 0.4, 0.5 \ \Gamma$ from the uppermost to lowermost curve.} 
\label{fig:spectral}
\end{figure}

 Most comprehensive choice of the frequency-dependent efficiencies is as follows:
 \begin{equation}
 \label{eq:two-wind}
 f_1=\Theta(\nu -\omega_s), f_2=\Theta(\omega_s - \nu).
 \end{equation}
The fist signal is thus collected from all  frequencies above the separating frequency $\omega_s$, while the second one comes from all frequencies below $\omega_s$.
The dimensionless  normalized cross-correlation $s_{12} \equiv S_{12} /\sqrt{\bar{I}_1 \bar{I}_2}$ is plotted in Fig. \ref{fig:spectral} versus $E$ at zero detuning and for several values of $\omega_s$. At low $E$, the correlations are formed by emission of photon pairs at opposite frequencies. At $\omega_s=0$, the numbers of photons emitted in two windows are precisely the same, this results in ideal shot-noise correlation $s_{12}=1$. At $\omega_s \ne 0$ only a part of the pairs are separated into different windows, so the correlation is smaller. The cross-correlation grows with increasing $E$ owing to photon bunching. At $\omega_s=0$, the cross-correlation diverges at the threshold. At $\omega_s \ne 0$, the growth changes to decrease and the normalized cross-correlation vanishes at the threshold. The reason for that is the narrowing of the spectral intensity upon approaching the threshold, so that the correlated emissions concentrate in one of the windows.  

\section{Conclusions}
\label{sec:conclusion}
To conclude, we have studied full counting statistics of Josephson junction circuit in the regime of parametric resonance. This is important in view of recent experiments that enable the detection of full power dissipated. We present the interpretation of statistics in terms of bursts of multiple-pairs of photons. We support this interpretation by investigating the time-dependent and frequency-resolved correlations. 

So far our results are restricted to the parameter region below the threshold where the field correlations are Gaussian. It is very interesting to address full counting statistics and time-dependent correlations in close vicinity of the instability threshold where the non-linear effects are important. This will be the subject of future research. 

\begin{acknowledgments}
The authors would like to thank
M. Hofheinz,  F. Portier and C.W.J. Beenakker for providing the motivation for this work and useful discussions. This research was supported by the Dutch Science
Foundation NWO/FOM.
\end{acknowledgments}

\appendix*
\section{Propagator}
For the perturbative calculations presented in the main text, we need the propagator of the fields $\varphi^\alpha$ or, equivalently, fields $\psi$, defined as 
\begin{equation}
\label{eq:propagator-definition-app}
G_{\alpha\beta} (t,t') = \langle \psi_\alpha^*(t) \psi_\beta(t')\rangle ,
\end{equation}
at $\chi=0$.
We rewrite the action at $\chi=0, n_\Omega=0$ with the aid of a dimensionless matrix $\ti{A}_{\nu}$
\begin{align}
S=&\ \frac{i}{32\pi G_Q Z_0}\frac{\Gamma^2}{4}\int\frac{d\ti{\nu}}{2\pi}\opcl{\psi_{\nu}^{\alpha}}^*\ti{A}_{\nu}^{\alpha\beta}\psi_{\nu}^{\beta};\\
\ti{A}_{\nu}=&\ 
\left( \begin{array}{cccc}
a(\ti{\nu},\ti{\nu}_0)\ & 0\ & iE\ & 0 \\
-2 \ & a(-\ti{\nu},-\ti{\nu}_0)\ & 0\ & -iE \\
iE \ & 0\ & a(-\ti{\nu},\ti{\nu}_0)\ & -2\\
0\ & -iE \ & 0\ & a(\ti{\nu},-\ti{\nu}_0)
 \end{array} \right)\notag\\
\ &\ a(x,y)= 1-i(x+y).\notag
\end{align}
The propagator in frequency domain is readily obtained by inverting $\ti{A}_{\nu}$.
\begin{align}
G(\nu)= 2^7 \pi G_Q Z_0 \Gamma^{-2}\ti{A}^{-1}(\nu).
\end{align}
The determinant of the action matrix 
\begin{align}
\det(\ti{A})= (\ti{\nu}^2+\gamma_+^2)(\ti{\nu}^2+\gamma_-^2)\ .
\end{align}
has four (generally complex) roots  at dimensionless frequencies $\pm i\gamma_{\pm}$, where $ \gamma_{\pm}=1\pm\sqrt{1-d}$.

The  propagator in time-domain is obtained by the inverse Fourier transform.
%\begin{align}
%\int dt\int dt'\opcl{\phi_{t}^{\alpha}}^*G_{t-t'}^{\alpha\beta}\phi_{t'}^{\beta} =&\ \frac{\Gamma}{2}\int\frac{d\ti{\nu}}{2\pi}\opcl{\phi_{\nu}^{\alpha}}^*G_{\nu}^{\alpha\beta}\phi_{\nu}^{\beta}
%\end{align}
%
We separate advanced $(t>t')$ and retarded $(t<t')$  part of the propagator. 
For advanced part,
\begin{align}
G_A=&\ G_A^+\ e^{-\gamma^{+} (t-t') \Gamma/2} + G^-_A\ e^{-\gamma^{-} (t-t') \Gamma/2}\ .
\end{align}
where $4\times 4$ matrices $G_A^{\pm}$ read:
\begin{align}
G_A^+=&\ \frac{16 \pi G_Q Z_0}{\gamma_+\sqrt{1-d}} 
\left( \begin{array}{cccc}
G_{11}\ & -E^2\ & G_{13}\ & G_{14} \\
G_{11} \ & -E^2\ & G_{13}\ & G_{14} \\
G_{14}^*\ & G_{13}^*\ & -E^2\ & G_{11}^* \\
G_{14}^*\ & G_{13}^*\ & -E^2\ & G_{11}^*
 \end{array} \right)\\
 &\text{with temporary notations} \notag \\
G_{11}=&\ E^2+2\opcl{1-i\ti{\nu}_0}\opcl{\sqrt{1-d}-i\ti{\nu}_0}\notag \\
G_{13}=&\ -iE\opcl{\sqrt{1-d}-i\ti{\nu}_0}\notag \\
G_{14}=&\ iE\opcl{2+i\ti{\nu}_0+\sqrt{1-d}}\notag
\end{align}

and 
\begin{align}
G_A^-=&\ \frac{16 \pi G_Q Z_0}{\gamma_-\sqrt{1-d}}\left( \begin{array}{cccc}
G_{11}\ & E^2\ & G_{13}\ & G_{14} \\
G_{11} \ & E^2\ & G_{13}\ & G_{14} \\
G_{14}^*\ & G_{13}^*\ & E^2\ & G_{11}^* \\
G_{14}^*\ & G_{13}^*\ & E^2\ & G_{11}^*
 \end{array} \right) \\
 &\text{with temporary notations} \notag \\
G_{11}=&\ -E^2+2\opcl{1-i\ti{\nu}_0}\opcl{\sqrt{1-d}+i\ti{\nu}_0}\notag \\
G_{13}=&\ -iE\opcl{\sqrt{1-d}+i\ti{\nu}_0}\notag \\
G_{14}=&\ -iE\opcl{2+i\ti{\nu}_0-\sqrt{1-d}}\notag
\end{align}
For the retarded part,
\begin{align}
G_R=&\ G_R^+\ e^{-\gamma^{+} (t'-t) \Gamma/2} + G^-_R\ e^{-\gamma^{-} (t'-t) \Gamma/2}\ .
\end{align}
where $4\times 4$ matrices $G_R^{\pm}$ read:
\begin{align}
G_R^+=&\ \frac{16 \pi G_Q Z_0}{\gamma_+\sqrt{1-d}}\left( \begin{array}{cccc}
-E^2\ & -E^2\ & G_{41}^*\ & G_{41}^* \\
G_{21} \ & G_{21} \ & G_{31}^*\ & G_{31}^* \\
G_{31}\ & G_{31}\ & G_{21}^* \ & G_{21}^*  \\
G_{41}\ & G_{41}\ & -E^2\ & -E^2
 \end{array} \right)\ ,\\
 &\text{with temporary notations} \notag \\ 
G_{21}=&\ E^2+2\opcl{1+i\ti{\nu}_0}\opcl{\sqrt{1-d}+i\ti{\nu}_0}\notag \\
G_{31}=&\ -iE\opcl{2-i\ti{\nu}_0+\sqrt{1-d}}\notag \\
G_{41}=&\ iE\opcl{\sqrt{1-d}+i\ti{\nu}_0}\notag
\end{align}
and
\begin{align}
G_R^-=&\ \frac{16 \pi G_Q Z_0 }{\gamma_-\sqrt{1-d}}\left( \begin{array}{cccc}
E^2\ & E^2\ &  G_{41}^*\ & G_{41}^* \\
G_{21} \ & G_{21} \ & G_{31}^*\ & G_{31}^* \\
G_{31}\ & G_{31}\ & G_{21}^* \ & G_{21}^*  \\
G_{41}\ & G_{41}\ & E^2\ & E^2
 \end{array} \right)\ ,\\
  &\text{with temporary notations} \notag \\
G_{21}=&\ -E^2+2\opcl{1+i\ti{\nu}_0}\opcl{\sqrt{1-d}-i\ti{\nu}_0}\notag \\
G_{31}=&\ iE\opcl{2-i\ti{\nu}_0-\sqrt{1-d}}\notag \\
G_{41}=&\ iE\opcl{\sqrt{1-d}-i\ti{\nu}_0}\notag
\end{align}

      \end{document}